\documentstyle[prb,preprint,aps]{revtex}

\draft

\begin{document}
\title{Specific heat and magnetic measurements in Nd$_{0.5}$Sr$_{0.5}$MnO$_{3}$, Nd$%
_{0.5}$Ca$_{0.5}$MnO$_{3}$ and Ho$_{0.5}$Ca$_{0.5}$MnO$_{3}$ samples}
\author{J. L\'{o}pez and O. F. de Lima}
\address{Instituto de F\'{i}sica Gleb Wataghin, Universidade Estadual de Campinas,\\
UNICAMP, 13083-970, Campinas, SP, Brazil}
\author{P. N. Lisboa-Filho and F. M. Araujo-Moreira}
\address{Depto. de F\'{i}sica, Universidade Federal de S\~{a}o Carlos, CP-676,
S\~{a}o Carlos, SP, 13565-905, Brazil}
\maketitle

\begin{abstract}
We studied the magnetization as a function of temperature and magnetic field
in the compounds Nd$_{0.5}$Sr$_{0.5}$MnO$_{3}$, Nd$_{0.5}$Ca$_{0.5}$MnO$_{3}$
and Ho$_{0.5}$Ca$_{0.5}$MnO$_{3}$. It allowed us to identify the
ferromagnetic, antiferromagnetic and charge ordering phases in each case.
The intrinsic magnetic moments of Nd$^{3+}$ and Ho$^{3+}$\ ions experienced
a short range order at low temperatures. We also did specific heat
measurements with applied magnetic fields between 0 and 9 T and temperatures
between 2 and 300 K in all three samples. Close to the charge ordering and
ferromagnetic transition temperatures the specific heat curves showed peaks
superposed to the characteristic response of the lattice oscillations. Below
10 K the specific heat measurements evidenced a Schottky-like anomaly for
all samples. However, we could not successfully fit the curves to either a
two level nor a distribution of two-level Schottky anomaly. Our results
indicated that the peak temperature of the Schottky anomaly was higher in
the compounds with narrower conduction band.
\end{abstract}


\section{Introduction}

Compounds like La$_{0.5}$Ca$_{0.5}$MnO$_{3}$\ and Nd$_{0.5}$Sr$_{0.5}$MnO$%
_{3}$\ present a real-space ordering of Mn$^{3+}$ and Mn$^{4+}$ ions, named
as charge ordering (CO). Close to the charge ordering temperature (T$_{CO}$%
), these materials show anomalies in resistivity, magnetization and lattice
parameters as a function of temperature, magnetic field and isotope mass\cite
{Radaelli}$^{,}$\cite{Zhao2}. At low temperatures both, ferromagnetic and
antiferromagnetic phases, could coexist \cite{Moritomo}. However, a
relatively small external magnetic field can destroy the CO phase and
enforces a ferromagnetic orientation of the spins\cite{Xiao}.

Moreover, electron microscope analysis has revealed convincing evidence that
CO is accompanied by orientational ordering of the 3d$^{3}$ orbitals on the
Mn$^{3+}$ ions, called orbital ordering (OO)\cite{Mori}. The physical
properties in CO manganese perovskites are thought to arise from the strong
competition among a ferromagnetic double exchange interaction, an
antiferromagnetic superexchange interaction, and the spin-phonon coupling.
These interactions are determined by intrinsic parameters such as doping
level, average cationic size, cationic disorder and oxygen stoichiometry.
Microscopically, CO compounds are particularly interesting because spin,
charge and orbital degrees of freedom are at play simultaneously and
classical simplifications that neglect some of these interactions do not
work. More detailed information on the physics of manganites can be found in
a review paper by Myron B. Salamon and Marcelo Jaime\cite{Myron}.

We have shown that polycrystalline samples of La$_{0.5}$Ca$_{0.5}$MnO$_{3}$
and Nd$_{0.5}$Sr$_{0.5}$MnO$_{3}$ presented an unusual magnetic relaxation
behavior close to each critical temperature\cite{J.López1}, \cite{J.López2}.
However, a clear understanding of all these features has not been reached
yet. An alternative to a bulk characterization like magnetization would be
to perform specific heat measurements. In contrast to magnetization, which
has a vector character, the specific heat is an scalar. Therefore, a
comparison between both types of data could give valuable information.

J. E. Gordon et al.\cite{Gordon} reported specific heat measurements for a Nd%
$_{0.67}$Sr$_{0.33}$MnO$_{3}$ sample and found a Schottky-like anomaly at
low temperatures. They associated this result to the magnetic ordering of Nd$%
^{3+}$ ions and the crystal-field splitting at low temperature. F.
Bartolom\'{e} et al.\cite{Bartolomé} also found Schottky-like anomaly in a
closely related compound of NdCrO$_{3}$. They proposed a crystal-field
energy level scheme in agreement with neutron-scattering studies in the same
sample. In two papers V. N. Smolyaninova et al.\cite{Smolyaninova1}, \cite
{Smolyaninova2} studied the low temperature specific heat in Pr$_{1-x}$Ca$%
_{x}$MnO$_{3}$ (0.3%
\mbox{$<$}%
x%
\mbox{$<$}%
0.5) and La$_{1-x}$Ca$_{x}$MnO$_{3}$ (x=0.47, 0.5 and 0.53). They found an
excess specific heat, C%
\'{}%
(T), of non-magnetic origin associated with charge ordering. They also
reported that a magnetic field sufficiently high to induce a transition from
the charge ordered state to the ferromagnetic metallic state did not
completely remove C%
\'{}%
(T). However, no Schottky anomaly was found in any of these compounds.

Here, we report magnetic and specific heat measurements with applied
magnetic fields between 0 and 9 T and temperatures between 2 and 300 K for Nd%
$_{0.5}$Sr$_{0.5}$MnO$_{3}$, Nd$_{0.5}$Ca$_{0.5}$MnO$_{3}$ and Ho$_{0.5}$Ca$%
_{0.5}$MnO$_{3}$ samples. All these compounds presented a Schottky-like
anomaly at low temperatures, as well as characteristic peaks associated to
the charge ordered, antiferromagnetic and/or ferromagnetic transitions. We
have already reported a short version of preliminary results about these
topics \cite{JLópez3}. However, as far as we know, detailed specific heat
measurements in these compounds have not been published.

\section{Experimental methods}

Polycrystalline samples of Nd$_{0.5}$Sr$_{0.5}$MnO$_{3}$, Nd$_{0.5}$Ca$%
_{0.5} $MnO$_{3}$ and Ho$_{0.5}$Ca$_{0.5}$MnO$_{3}$ were prepared by the
sol-gel method\cite{Paulo}. Stoichiometric parts of Nd$_{2}$O$_{3}$ (Ho$_{2}$%
O$_{3}$) and MnCO$_{3}$ were dissolved in HNO$_{3}$ and mixed to an aqueous
citric acid solution, to which SrCO$_{3}$ or CaCO$_{3}$\ was added. The
mixed metallic citrate solution presented the ratio citric acid/metal of 1/3
(in molar basis). Ethylene glycol was added to this solution, to obtain a
citric acid/ethylene glycol ratio 60/40 (mass ratio). The resulting solution
was neutralized to pH$\sim $7 with ethylenediamine. This solution was turned
into a gel, and subsequently decomposed to a solid by heating at 400 $^{o}$%
C. The resulting powder was heat-treated in vacuum at 900 $^{o}$C for 24
hours, with several intermediary grindings, in order to prevent formation of
impurity phases. This powder was pressed into pellets and sintered in air at
1050 $^{o}$C for 12 hours. X-ray diffraction measurements indicated high
quality samples in all cases.

The magnetization measurements were done with a Quantum Design MPMS-5S SQUID
magnetometer. Specific heat measurements were done with a Quantum Design
PPMS calorimeter. The PPMS used the two relaxation time technique, and data
was always collected during sample cooling. The intensity of the heat pulse
was calculated to produce a variation in the temperature bath between 0.5 \%
(at low temperatures) and 2\% (at high temperatures). Experimental errors
during the specific heat and magnetization measurements were lower than 1 \%
for all temperatures and samples.

\section{Results and Discussion}

\subsection{Magnetization measurements}

Figure 1 shows the temperature dependence of magnetization, measured with a
5 T applied magnetic field and zero field cooling conditions, in
polycrystalline samples of Nd$_{0.5}$Sr$_{0.5}$MnO$_{3}$, Nd$_{0.5}$Ca$%
_{0.5} $MnO$_{3}$ and Ho$_{0.5}$Ca$_{0.5}$MnO$_{3}$. All three compounds
present a charge ordering transition below 160, 250 and 271 K, respectively 
\cite{Kajimoto}$^{,}$ \cite{Millange}$^{,}$ \cite{Terai}. It is interesting
to note that the relation between the charge ordering temperature and the
antiferromagnetic ordering temperature changes from one sample to the other.
In the first case they are approximately coincident (T$_{CO}\approx $T$_{N}$%
), in the second case the charge ordering temperature is much higher (T$%
_{CO}\gg $T$_{N}$), and in the third case there is not a long range
antiferromagnetic transition.

The Nd$_{0.5}$Sr$_{0.5}$MnO$_{3}$ sample presents a ferromagnetic transition
at T$_{C}$=250 K and an antiferromagnetic transition at T$_{N}$=160 K. The Nd%
$_{0.5}$Ca$_{0.5}$MnO$_{3}$ compound presents a strong maximum near T$_{CO}$%
, but shows an unexpected minimum close to the antiferromagnetic transition
temperature T$_{N}$=160 K. Usually an antiferromagnetic transition is
accompanied by a maximum in the temperature dependence of the magnetization.
For temperatures lower than 20 K both Nd$_{0.5}$Sr$_{0.5}$MnO$_{3}$ and Nd$%
_{0.5}$Ca$_{0.5}$MnO$_{3}$ samples show a sharp increase in the
magnetization. This sharp increase in magnetization have been associated to
a short range magnetic ordering of the intrinsic magnetic moment of Nd$^{3+}$
ions\cite{Mathieu}. However, no long range ferromagnetic order of the Nd$%
^{3+}$ ions was found in neutron diffraction measurements at these low
temperatures\cite{Kajimoto}$^{,}$ \cite{Millange}.

Differently from the two previous samples, the Ho$_{0.5}$Ca$_{0.5}$MnO$_{3}$
compound do not present a strong maximum at the charge ordering temperature
in the magnetization versus temperature curve. The existence of charge
ordering in Ho$_{0.5}$Ca$_{0.5}$MnO$_{3}$ was suggested by\ T. Terai et al. 
\cite{Terai} studying simultaneously the magnetization and resistivity
curves. Figure 2 shows the inverse of the DC susceptibility measured with
0.1 mT (inset) and 5 T (main figure) in the Ho$_{0.5}$Ca$_{0.5}$MnO$_{3}$\
sample. The measurement at 5 T clearly indicates a transition between two
linear behaviors near 270 K. This latter result, together with the specific
heat measurements, that we will discuss below, reinforces the idea of the
existence of a charge ordering transition in the Ho$_{0.5}$Ca$_{0.5}$MnO$%
_{3} $ sample, although the details of the interactions in this compound
differ from those in Nd$_{0.5}$Sr$_{0.5}$MnO$_{3}$ and Nd$_{0.5}$Ca$_{0.5}$%
MnO$_{3}$. On the other hand, the DC susceptibility measurement in Ho$_{0.5}$%
Ca$_{0.5} $MnO$_{3}$ with a probe field of 0.1 mT indicates two possible
Curie-Weiss intervals: below and above 100 K (see straight lines).

We fitted the magnetization data, measured with a probe field of 0.1 mT for
all samples, to a Curie-Weiss law: $M/H\sim \mu _{eff}^{2}/(T-T_{\Theta })$,
where $M/H$ was the DC susceptibility, $\mu _{eff}$\ was the effective
paramagnetic moment and $T_{\Theta }$\ was the Curie-Weiss temperature. A
simplified theoretical estimate\cite{Millange} of the effective paramagnetic
moment gives values of 5.1 $\mu _{B}$, 5.1 $\mu _{B}$ and 8.7 $\mu _{B}$ for
Nd$_{0.5}$Sr$_{0.5}$MnO$_{3}$, Nd$_{0.5}$Ca$_{0.5}$MnO$_{3}$ and Ho$_{0.5}$Ca%
$_{0.5}$MnO$_{3}$ samples, respectively. The positive or negative value of $%
T_{\Theta }$ suggests the predominance of local ferromagnetic or
antiferromagnetic interactions between the spins in the corresponding
interval of temperature used. However, it is important to remember that the
Curie-Weiss law is only a mean field approximation, that do not consider
fluctuations.

In our case the fitting interval was taken between 300 and 400 K in the Nd$%
_{0.5}$Sr$_{0.5}$MnO$_{3}$ and Nd$_{0.5}$Ca$_{0.5}$MnO$_{3}$ samples. The $%
\mu _{eff}$ and $T_{\Theta }$ values were 2.2 $\mu _{B}$ and 2.1 $\mu _{B}$,
and 260 K and 215 K for Nd$_{0.5}$Sr$_{0.5}$MnO$_{3}$ and Nd$_{0.5}$Ca$%
_{0.5} $MnO$_{3}$ samples, respectively. The discrepancies between our
experimentally measured values of $\mu _{eff}$\ and the theoretical ones in
the samples of Nd$_{0.5}$Sr$_{0.5}$MnO$_{3}$ and Nd$_{0.5}$Ca$_{0.5}$MnO$%
_{3} $, are essentially due to the vicinity of the T$_{C}$ or T$_{CO}$
temperatures to the temperature interval that we could use to fit the curves
to the Curie-Weiss law. F. Millange et al. \cite{Millange} measured the
effective paramagnetic moment, in the interval between 450 and 800 K, for a
Nd$_{0.5}$Ca$_{0.5}$MnO$_{3}$ sample and found a value roughly in agreement
with theory. M. T. Causa et al.\cite{Causa} also showed that the transition
to a Curie-Weiss regime in several manganites occurs approximately for T 
\mbox{$>$}%
2\thinspace T$_{C}$ . In the temperature interval approximately between 2$\,$%
T$_{C}$ and T$_{C}$ magnetic ions start to form local clusters. This
behavior is different from the one of a simple ferromagnetic sample like
iron, where the transition from the ferromagnetic state to the Curie-Weiss
regime occurs in a much shorter temperature interval.

For the Ho$_{0.5}$Ca$_{0.5}$MnO$_{3}$ sample we fitted the Curie-Weiss law
to two intervals: from 100 to 350 K and from 2 to 100 K. In the Ho$_{0.5}$Ca$%
_{0.5}$MnO$_{3}$\ sample the $\mu _{eff}$ and $T_{\Theta }$ values were 9.7 $%
\mu _{B}$ and 3.4 $\mu _{B}$, and -38 K and -16 K for the high and low
temperature intervals, respectively. The effective paramagnetic moment found
in the high temperature interval for the Ho$_{0.5}$Ca$_{0.5}$MnO$_{3}$\
sample is roughly in agreement with the theoretical value, and the negative
value of $T_{\Theta }$ indicates that the main local interactions are
antiferromagnetic. However, the transition approximately below 100 K to a
different Curie-Weiss regime suggests that the predominance of short range
antiferromagnetic interactions are giving space to short range ferromagnetic
ones, with a higher value of $T_{\Theta }$. This conclusion is also
supported by the magnetic field dependence of the magnetization, as we will
discuss below.

Figure 3 shows the magnetization hysteresis cycle at 2 K for the three
studied samples. The applied magnetic field was increased from 0 to 5 T,
decreased to -5 T and then increased back to 5 T again. The Nd$_{0.5}$Sr$%
_{0.5}$MnO$_{3}$\ curve is characteristic of a mixture of two phases: one
ferromagnetic and another antiferromagnetic. The ferromagnetic part is
easily oriented at low magnetic field values and shows a hysteretical
behavior. The almost linear and reversible dependence for magnetic fields
higher than approximately 1 T, indicates a gradual destruction of the
antiferromagnetic phase\cite{J.López2}. R. Mahendiram et al.\cite{Mahendiran}
reported, in a sample with the same composition at 50 K, that for magnetic
fields higher than approximately 5 T, the magnetization started to increase
rapidly, and for magnetic fields above 10 T, it slowly approached the
ferromagnetic saturation value. Because at 2 K these transition fields are
much higher than 5 T we were unable to see them. The Nd$_{0.5}$Ca$_{0.5}$MnO$%
_{3}$\ curve did not show any trace of a ferromagnetic component at low
fields. In fact, the curve is linear and reversible for the whole magnetic
field interval. As before, this linearity characterizes the gradual
destruction of the antiferromagnetic phase. F. Millange et al.\cite{Millange}
reported, for a sample with the same composition, a set of M vs. H curves
with applied magnetic fields up to 22 T. They found sharp transitions and a
large hysteresis at 130 K, between applied magnetic fields of 12 and 18 T.
Their results were interpreted as evidence of the existence of a spin-flop
transition.

In the M vs. H curve for the Ho$_{0.5}$Ca$_{0.5}$MnO$_{3}$ sample, two
characteristics are well noticed. One, that there is not hysteresis at all,
and second, that the magnetization values at 5 T are well below the
theoretical saturation value of 8.7 $\mu _{B}$. In order to try to
discriminate between a paramagnetic or ferromagnetic ordering at 2 K, the
experimental points were fitted by a Brillouin function\cite{Ashcroft}:

\begin{equation}
M=(N/V)\,\gamma \,J\,B_{J}(x)  \label{1}
\end{equation}
where $\gamma =g\,\mu _{B},\,\,\beta =1/k_{B}T$ and

\begin{equation}
B_{J}(x)=[(2J+1)/2J]\coth [(2J+1)x/2J]-(1/2J)\coth (x/2J)  \label{2}
\end{equation}

Here, $x=\beta \,\gamma \,J\,H$, $N/V{\em \ }$is the number of ions per unit
volume, $g$ is an effective gyromagnetic ratio,{\em \ }${\em J}$ is the
total angular momentum,${\em T}$ is the temperature, $\mu _{B}$\ is the Bohr
magneton and $k_{B}$\ is the Boltzmann constant. Equation 1 is valid for a
set of identical and not interacting ions of angular momentum {\em J}.
Because our sample has three different magnetic ions we fixed J=8 (Ho$^{3+}$%
), J=2 (Mn$^{3+}$) or J=3/2 (Mn$^{4+}$) and let the effective gyromagnetic
ratio to change freely in each case. The best fitting, represented by the
solid line in figure 3, was found for {\em J=8} and {\em g=0.35}. Although
it is difficult to enunciate a more conclusive statement from these results,
they suggest that there is not a long range ferromagnetic order for the Ho$%
_{0.5}$Ca$_{0.5}$MnO$_{3}$ sample, at 2 K.

\subsection{Specific heat at high temperatures}

Figure 4 shows specific heat measurements with a zero applied magnetic field
from 2 to 300 K in the Nd$_{0.5}$Sr$_{0.5}$MnO$_{3}$, Nd$_{0.5}$Ca$_{0.5}$MnO%
$_{3}$ and Ho$_{0.5}$Ca$_{0.5}$MnO$_{3}$ samples. In order to facilitate the
visualization, the curves for Nd$_{0.5}$Sr$_{0.5}$MnO$_{3}$ and Ho$_{0.5}$Ca$%
_{0.5}$MnO$_{3}$ were displaced 20 J/mol K upside and downside,
respectively. Specific heat measurements give information about both lattice
and magnetic excitations. At high temperatures the excitations from the
lattice vibrations are dominant and decrease as the temperature decreases.
The magnetic contribution can be obtained by subtracting the lattice part
from the experimental values.

The Dulong and Petit\cite{Kittel} model predicts that the thermal
contribution at high temperature increases assintotically to the value $%
C_{DP}=3nR$ , where $n$ is the number of atoms in the compound%
\'{}%
s formulae and $R$ is the ideal gas constant. Therefore, $C_{DP}=$125 J/(mol
K) for all studied samples in this work, since their unit cell contain 5
atoms each. At 300 K the experimental specific heat values were 110 J/(mol
K) (88 \% of $C_{DP}$) for the Nd$_{0.5}$Sr$_{0.5}$MnO$_{3}$ and Nd$_{0.5}$Ca%
$_{0.5}$MnO$_{3}$ samples and 100 J/(mol K) (80 \% of $C_{DP}$) for the Ho$%
_{0.5}$Ca$_{0.5}$MnO$_{3}$ sample. These values suggest that the Debye
temperature should be, whithin the experimental error, approximately equal
for all three compounds.

Continuous lines in figure 4 represent the fitting of the thermal
background, in the interval from 30 to 300 K, by the Einstein model given by:

\begin{equation}
C_{Einstein}=3nR\sum_{i}a_{i}\left[ \frac{x_{i}^{2}\,e^{x_{i}}}{\left(
e^{x_{i}}-1\right) ^{2}}\right]  \label{3}
\end{equation}
where $x_{i}=T_{i}/T$ . We used three optical phonons ($i$ =1, 2, 3) with
energies ${\em T}_{i}$ (in Kelvin) and relative occupations ${\em a}_{i}$.
The Einstein model for the specific heat considers the oscillation frequency
(or energy) independently from the wave vector, which is a valid
approximation for the optical part of the spectrum. The values of
temperatures (energies) found were 148, 438 and 997 K for Nd$_{0.5}$Sr$%
_{0.5} $MnO$_{3}$, 152, 432 and 1035 K for Nd$_{0.5}$Ca$_{0.5}$MnO$_{3}$ and
147, 438 and 1023 K for Ho$_{0.5}$Ca$_{0.5}$MnO$_{3}$. These values are
similar to those reported, using the same model, by A. P. Ramirez et al.\cite
{Ramirez} in a La$_{0.37}$Ca$_{0.63}$MnO$_{3}$\ sample and Raychaudhuri et
al.\cite{Raychaudhuri} in a Pr$_{0.63}$Ca$_{0.37}$MnO$_{3}$ sample. However,
we should point out that the values found are not unique because there are
six varying parameters during the fitting (one energy and one occupation
coefficient for each oscillation mode).

An alternative method to determine the oscillation mode energies of the
crystalline lattice is to study the Raman spectrum of a given compound. For
example, V. Dediu et al.\cite{Dediu} found that, at a measured temperature
of 50 K for a Pr$_{0.65}$Ca$_{0.35}$MnO$_{3}$ sample, the strongest peaks
were located at energies of 683 and 877 K. Besides, E. Granado et al.\cite
{Granado} found that, at a measured temperature of 200 K for a Nd$_{0.66}$Ca$%
_{0.33}$MnO$_{3}$ sample, the strongest peaks were located at energies of
388, 439, 690 and 877 K. However, these peaks were not the only ones, they
could change with temperature and they are not narrow. Furthermore, the
Raman spectrum could depend on variables like the frequency, the intensity
of the excitation laser and the luminosity of the sample. All of these make
difficult a straightforward comparison between both methods to determine the
oscillation energies of a crystalline lattice.

Figure 5 represents the differences between the experimental data and the
fitted curves in figure 4. There is a maximum at 231 K for the Nd$_{0.5}$Sr$%
_{0.5}$MnO$_{3}$ sample, which is correlated with the ferromagnetic
transition at 250 K in the corresponding magnetization curve (figure 1). A
second maximum, which could be partially associated to the antiferromagnetic
and charge ordering transitions at 160 K, appears at 180 K. However, lattice
parameters in this compound change rapidly between approximately 110 K and
250 K\cite{Shimomura}. This variation in lattice parameters changes the
intensity of the interactions between the atoms, and consequently the
oscillation frequency of the phonons mode, contributing to the specific heat
in the second peak. Figure 5 also shows that there is a maximum at 243 K for
the Nd$_{0.5}$Ca$_{0.5}$MnO$_{3}$ sample. This maximum correlates with the
corresponding charge ordering temperature at 250 K. Differently from the Nd$%
_{0.5}$Sr$_{0.5}$MnO$_{3}$ sample, there is not magnetic ordering in this
high temperature interval for the Nd$_{0.5}$Ca$_{0.5}$MnO$_{3}$ sample. In
this latter compound lattice parameters change very rapidly between
approximately 200 and 250 K \cite{Millange}. An inflection point in the C
vs. T curve appears at 141 K for the Nd$_{0.5}$Ca$_{0.5}$MnO$_{3}$ sample.
Besides, there is a maximum at 141 K in figure 5, but its height is
relatively small compared to the maximum at 243 K. The Neel temperature
corresponding to this compound is 160 K. Therefore, these results lead us to
conclude that the specific heat variations due to the antiferromagnetic
order are small compared to those induced in the charge ordering and
ferromagnetic transitions. For the Ho$_{0.5}$Ca$_{0.5}$MnO$_{3}$ sample, we
found a maximum at 276 K. This maximum is correlated with the transition
seen at high magnetic fields in the inverse DC susceptibility curve of
figure 2. These experimental results and the resistivity measurements
reported by T. Terai et al.\cite{Terai}, indicate the existence of a phase
transition between 270 and 280 K. However, electron diffraction studies
would be needed to unambiguously classify this transition as CO.

V. Hardy et al.\cite{Hardy} reported similar measurements in a single
crystal of Pr$_{0.63}$Ca$_{0.37}$MnO$_{3}$. They found a specific heat peak
at 229 K and resistivity and magnetization curves indicated a charge
ordering transition at 227 and 228 K, respectively. They also point out that
part of the peak amplitude was due to the crystalline lattice distortions.
Moreover, the specific heat curve showed an inflection point at 150 K, close
to the corresponding T$_{N}$ = 160 K. Similar characteristics were also
found in a polycrystalline sample of La$_{0.37}$Ca$_{0.63}$MnO$_{3}$\cite
{Ramirez}.

Results in figure 5 allow us to calculate the variation in entropy ($\Delta $%
S), associated to the charge ordering, ferromagnetic and antiferromagnetic
transitions:

\begin{equation}
\Delta S=\int\limits_{Ti}^{Tf}\frac{\left( C-C_{ph}\right) }{T}\,dT
\label{4}
\end{equation}
where ${\em T}_{i}$ and ${\em T}_{f}$ are two temperatures conveniently
chosen to delimitate the interval of interest and ${\em C}_{ph}$ is the
specific heat due to the lattice oscillations.

For the Nd$_{0.5}$Ca$_{0.5}$MnO$_{3}$ sample the entropy variation between
201 and 301 K was $\Delta S($T$_{CO})=$ 2.0 J/(mol K). A. K. Raychaudhuri et
al.\cite{Raychaudhuri} reported an entropy variation, close to the charge
ordering transition in the compound Pr$_{0.63}$Ca$_{0.37}$MnO$_{3}$, of 1.8
J/(mol K) with zero applied magnetic field and 1.5 J/(mol K) with an 8 T
magnetic field. On the other hand, Ramirez et al. \cite{Ramirez} found $%
\Delta S($T$_{CO})=$ 5 J/(mol K) in a La$_{0.37}$Ca$_{0.63}$MnO$_{3}$
sample. All these results correspond to those expected for a charge ordering
transition \cite{Raychaudhuri}. On the other hand, the entropy variation
(not related with phonons), calculated between 118 and 201 K for the Nd$%
_{0.5}$Ca$_{0.5}$MnO$_{3}$ sample, was $\Delta S($T$_{N})=$ 0.80 J/(mol K).
We associate this smaller $\Delta {\em S}$\ value to the antiferromagnetic
order at T$_{N}=$160 K.

The entropy variation between 133 and 274 K for the Nd$_{0.5}$Sr$_{0.5}$MnO$%
_{3}$ sample was $\Delta S($T$_{CO+FM})=$ 3.6 J/(mol K). Considering that
the entropy variation associated to the charge ordering transition is the
same as for the Nd$_{0.5}$Ca$_{0.5}$MnO$_{3}$ sample, we found that the
entropy variation associated to the ferromagnetic transition is
approximately 1.6 J/(mol K). Using a model proposed by J. E. Gordon et al. 
\cite{Gordon} we estimated that the change in entropy needed to a full
ferromagnetic transition in the Nd$_{0.5}$Sr$_{0.5}$MnO$_{3}$ sample was
12.45 J/(mol K). Therefore, the fact that the entropy variation found is 13
\% of the theoretical value, confirms that only a small part of the spins
order ferromagnetically. The same report of J. E. Gordon et al. \cite{Gordon}
found that the entropy variation associated to the ferromagnetic order in a
Nd$_{0.67}$Sr$_{0.33}$MnO$_{3}$ sample was approximately 10 \% of the
theoretical value. The entropy variation, not associated to lattice
oscillations, between 216 and 294 K for the Ho$_{0.5}$Ca$_{0.5}$MnO$_{3}$
sample was $\Delta S($T$_{CO})=$ 0.78 J/(mol K). This $\Delta S($T$_{CO})$\
for the Ho$_{0.5}$Ca$_{0.5}$MnO$_{3}$\ sample is smaller than the
corresponding value for the Nd$_{0.5}$Ca$_{0.5}$MnO$_{3}$ sample, suggesting
a different nature of the charge ordering transition.

\subsection{Specific heat at low temperatures}

For temperatures lower than 30 K, the phonon contribution to the specific
heat starts to take a smaller part, and the C vs. T curve shows a slow
variation. Figure 6 reproduces the specific heat measurements for
temperatures between 2 and 30 K in the samples of (a) Nd$_{0.5}$Sr$_{0.5}$MnO%
$_{3}$, (b) Nd$_{0.5}$Ca$_{0.5}$MnO$_{3}$ and (c) Ho$_{0.5}$Ca$_{0.5}$MnO$%
_{3}$. Measurements were made in the presence of applied magnetic fields of
0 T (open squares), 5 T (closed circles), 7 T (open up triangles) and 9 T
(closed down triangles). Note that close to 5 K all curves show a
Schottky-like anomaly\cite{Kittel}.

It is important to stress that high values of specific heat were found at
low temperatures for the three samples presented here. These values are
similar to those reported by J. E. Gordon et. al.\cite{Gordon} in a sample
of Nd$_{0.67}$Sr$_{0.33}$MnO$_{3}$. However, the absolute values of specific
heat at 2 K, reported by J. J. Hamilton et al.\cite{Hamilton} in samples of
La$_{0.67}$Ba$_{0.33}$MnO$_{3}$ and La$_{0.80}$Ca$_{0.20}$MnO$_{3}$, were
more than 100 times smaller. Similarly, V. Hardy et al.\cite{Hardy}, in a
single crystal of Pr$_{0.63}$Ca$_{0.37}$MnO$_{3}$, found values
approximately equal to those reported by J. J. Hamilton et al.\cite{Hamilton}%
. The high values of specific heat found in our work could be interpreted as
an increase in the effective mass of the electrons due to localization,
which is also consistent with the insulating behavior revealed by electrical
resistivity measurements\cite{Kajimoto}$^{,}$ \cite{Millange}$^{,}$ \cite
{Terai}.

Continuous lines in figure 6 indicate the fitting of the experimental data
between 15 and 30 K by the following expression\cite{Gordon}:

\begin{equation}
C=\sum \beta _{2n+1}\,T^{\text{\/}2n+1}  \label{5}
\end{equation}
Here, ${\em C}$ is the specific heat, ${\em T}$ is the temperature and the
parameters $\beta _{2n+1}$ represent the contribution of the phonon modes.
Notice that we did not include the lowest temperature interval to avoid the
Schottky anomaly. To be able to fit the whole temperature interval we have
chosen values of $n$ from 1 to 4. Nonetheless, this large number of\ free
parameters difficult an unique determination of each one. Since from
resistivity measurements\cite{Kawano}$^{,}$ \cite{Millange}$^{,}$ \cite
{Terai} all the studied samples show an insulating behavior at low
temperatures, and the applied magnetic fields are not strong enough to
destroy this characteristic, the expected linear contribution, from the free
electrons to the specific heat, is zero. However, other kind of excitations
could lead to a linear contribution. This can imply an implicit error of the
fitting model. Moreover, we could not resolve in our data a term of type T$%
^{3/2}$. This term is usually interpreted as an evidence of the existence of
ferromagnetic interactions. However, previous studies in samples clearly
identified by other techniques as ferromagnetic, have not found this term in
the specific heat either\cite{Smolyaninova2}$^{,}$ \cite{Hamilton}.

The values of $\beta _{3}$ change between 0.28 mJ/(mol K$^{4}$) at H=0 T in
Nd$_{0.5}$Sr$_{0.5}$MnO$_{3}$ to 1.57 mJ/(mol K$^{4}$) at H=5 T in Ho$_{0.5}$%
Ca$_{0.5}$MnO$_{3}$. The corresponding Debye temperatures (T$_{D}$),
obtained from $\beta _{3}$, will be plotted in figure 7a. The graphs in the
insets of figure 6 show the differences between the specific heat
experimental data and the phonon contribution to the specific heat,
extrapolated to low temperatures from the fitting in the temperature
interval between 15 and 30 K.

Figure 7 shows (a) the magnetic field dependence of the Debye temperature,
(b) the variation of magnetic entropy between 2 and 20 K ($\Delta S$) and
(c) the Schottky temperature (T$_{S}$) in the three studied samples. The
entropy variation was calculated from equation 4 and the Schottky
temperature was determined from the maxima\bigskip\ in the insets of figure
6. The Debye temperature was calculated using the values of $\beta _{3}$\
and the following equation\cite{Kittel}:

\begin{equation}
T_{D}=\left( \frac{12\pi ^{4}nR}{5\beta _{3}}\right) ^{1/3}  \label{6}
\end{equation}
where ${\em n}$ is the number of atoms in the unit cell and ${\em R}$ is the
ideal gas constant. We should point out that the Debye temperature was
estimated from the low temperature data. This procedure lead to a Debye
temperature that is slightly different than the actual value of temperature
for which the specific heat saturates. The Debye temperature from the low
temperature data decreases with the increase of the applied magnetic field.
We have also made specific heat measurements with a 9 T magnetic field, at
high temperatures, for several charge ordered compounds\cite{JLópez4}, and
they show an increase of the specific heat when compared to the 0 T case, in
agreement with the magnetic field dependence of the Debye temperatures.
Other authors\cite{Gordon} have made an initial assumption that the Debye
temperature is magnetic field independent, which is not supported by our
experimental results.

Figure 7 also shows that T$_{S}$ grows with the increase of the external
magnetic field in all cases. However, the growth of T$_{S}$ seems to be
saturated for a magnetic field of 5 T in the sample of Ho$_{0.5}$Ca$_{0.5}$%
MnO$_{3}$, while there is no sign of T$_{S}$ saturation\ in the other two
compounds. It is also interesting to note here the relative low T$_{S}$
values. For the reagent compounds of Nd$_{2}$O$_{3}$ and Ho$_{2}$O$_{3}$ the
peak in the specific heat, measured in zero magnetic field, were found at
approximately 10 and 9 K, respectively\cite{Touloukian}. The fact that T$%
_{S} $\ is lower in the manganese compounds suggests that the collective
charge ordered phase could be determining a smaller splitting in the energy
levels.

The entropy variation, associated to the Schottky anomaly, grows as a
function of magnetic field in the sample of Nd$_{0.5}$Sr$_{0.5}$MnO$_{3}$.
The same result is clearly visualized from the height of the Schottky
anomaly in the inset of figure 6a. J. E. Gordon et al.\cite{Gordon} also
reported a similar increase in a sample of Nd$_{0.67}$Sr$_{0.33}$MnO$_{3}$.
However, the Schottky entropy variation decreases with the increase of
magnetic field in the Nd$_{0.5}$Ca$_{0.5}$MnO$_{3}$ and Ho$_{0.5}$Ca$_{0.5}$%
MnO$_{3}$ samples. In these two cases, the C vs. T curves in figure 6b and
6c, indicate that the local minimum, at a temperature above the Schottky
anomaly, disappears with the increase of the external magnetic field. This
is also reflected in the decrease of the height of the peak with the
increase of the applied magnetic field (insets of figures 6b and 6c). These
results seem to indicate that the magnetic field dependence of the Schottky
anomaly is more correlated with the presence (Nd$_{0.5}$Sr$_{0.5}$MnO$_{3}$)
or not (Nd$_{0.5}$Ca$_{0.5}$MnO$_{3}$ and Ho$_{0.5}$Ca$_{0.5}$MnO$_{3}$) of
ferromagnetic interactions than with the particular type of magnetic ion (Nd$%
^{3+}$ or Ho$^{3+}$).

The expected entropy variation from the magnetic ordering of Nd$^{3+}$ or Ho$%
^{3+}$ ions could be estimated\cite{Gordon} as $\Delta S$=0.5 R ln(2), where 
${\em R}$ is the ideal gas constant. The actually found variation correspond
to values from 63 to 77 \% of the expected ones in the Nd$_{0.5}$Sr$_{0.5}$%
MnO$_{3}$ sample, 80 to 62 \% in the Nd$_{0.5}$Ca$_{0.5}$MnO$_{3}$ sample
and 68 to 25 \% in the Ho$_{0.5}$Ca$_{0.5}$MnO$_{3}$ sample, for magnetic
fields between 0 and 9 T, respectively. J. E. Gordon et al.\cite{Gordon}
found that the entropy variation associated to the ordering of Nd$^{3+}$
ions\ in Nd$_{0.67}$Sr$_{0.33}$MnO$_{3}$ was approximately 85 \% of the
expected value.

Let us consider that Nd (Ho) ions be oriented by a molecular field
interaction (H$_{mf}$), and not by the exchange interaction between pairs of
Nd-Nd ions (Ho-Ho)\cite{Gordon}. Assuming that H$_{mf}$\ does not change
with the external magnetic field, it is possible to estimate it, using a
mean field model and the peak temperature in the specific heat. Considering
a two level energy splitting ${\em \Delta (H)}$, due to a magnetic moment $%
{\em m}${\em \ }in an external magnetic field ${\em H}$, one finds in zero
applied field $\Delta (0)=2mH_{mf}$, and in ${\em H=9}$ T the value changes
to $\Delta (9$ T$)=2m[H_{mf}+9$ T$]$. One can also use that the energy
splitting could be related to the peak temperature in the specific heat by $%
\Delta {\em \ =k}_{B}{\em \ T}_{S}{\em /0.418}$, a relation valid for a two
level Schottky function\cite{Kittel}. Solving this system of two linear
equations we found that ${\em H}_{mf}{\em \ =11.4}$ T and ${\em m=0.43\,}\mu
_{B}$ \ in Nd$_{0.5}$Sr$_{0.5}$MnO$_{3}$, ${\em H}_{mf}{\em \ =20.6}$ T and $%
{\em m=0.44\,}\mu _{B}$ in Nd$_{0.5}$Ca$_{0.5}$MnO$_{3}$ and ${\em H}_{mf}%
{\em \ =13.6}$ T and ${\em m=0.58\,}\mu _{B}$ in Ho$_{0.5}$Ca$_{0.5}$MnO$_{3}
$.\ These values of ${\em m}$ are smaller than those obtained from
susceptibility measurements at high temperatures. However, they are similar
to the one found by J. E. Gordon et al.\cite{Gordon} using the same method ($%
H_{mf}=$10 T and $m=$ 0.8 $\mu _{B}$ in Nd$_{0.67}$Sr$_{0.33}$MnO$_{3}$).

The ground state of the Nd$^{3+}$ and Ho$^{3+}$ ions are usually denoted as $%
^{4}$I$_{9/2}$ and $^{5}$I$_{8}$, where I{\em \ }stands for an orbital
angular momentum L=6, the superprefix specify the total spin as ${\em 2S+1}$
and the subscript the total angular momentum ${\em J}$. The number of the
lowest energy levels is given by ${\em 2J+1}$, which leads to 5 Kramers
doublets in the ground state of the first ion and a singlet and 8 Kramers
doublets in the second ion\cite{Ashcroft}. F. Bartolom\'{e} et. al.\cite
{Bartolomé} showed that the second doublet in the Nd$^{3+}$ ion was
approximately 120 K (in energy) above the lowest doublet. As this
temperature is about 10 times higher than the temperature where the Schottky
anomaly appears, the contribution of the second doublet is expected to be
small. In a previous report\cite{JLópez3} we showed that a two level
Schottky function (only one doublet) did not fit properly our experimental
data at low temperatures. The same result was verified for the new
experimental data presented here. One alternative, justified by the
existence of several different grains in polycrystalline samples, is to
consider a distribution of energy splitting around the value that would
correspond to a single crystal in the same two level Schottky model.
Although the fitting results using this second approach improved a little
bit, we found that they still remained unsatisfactory.

At first sight someone might be tempted to correlate the existence of the
Schottky anomaly with the presence of an intrinsic magnetic moment in Nd$%
^{3+}$ and Ho$^{3+}$ ions (in contrast with La$^{3+}$ ions without magnetic
moment and no Schottky anomaly in the manganite). However, specific heat
measurements reported by V. Hardy et al.\cite{Hardy} in a compound of Pr$%
_{0.63}$Ca$_{0.37}$MnO$_{3}$ (Pr$^{3+}$ ions have approximately the same
magnetic moment as Nd$^{3+}$ ions) did not show any Schottky anomaly.
Moreover, Ho$^{3+}$ ions have an intrinsic magnetic moment almost 3 times
bigger than Nd$^{3+}$ ions, but the Schottky temperature at zero magnetic
field were 2.73 K in Nd$_{0.5}$Sr$_{0.5}$MnO$_{3}$, 5.08 K in Nd$_{0.5}$Ca$%
_{0.5}$MnO$_{3}$, and 4.39 K in Ho$_{0.5}$Ca$_{0.5}$MnO$_{3}$. Probably the
existence of the Schottky anomaly is related with the Kramers theorem. It
states that an ion possessing an odd number of electrons, no matter how
unsymmetrical the crystal field, must have a ground state that is at least
doubly degenerate\cite{Ashcroft}. This could lead to the thermal
depopulation that produces the Schottky anomaly in the specific heat. Ions
of Ce, Nd, Sm, Gd, Dy, Er and Yb all have an odd number of electrons and
their respective oxides present a Schottky anomaly in the specific heat.
However, the Kramers theorem does not exclude that ions with an even number
of electrons might also have a doubly degenerate ground state. This is the
case of the Ho ions, although it is noteworthy that the magnetic entropy
associated with the Schottky transition in the compound with Ho had the
lowest value (see above). Besides, our results indicate that the higher T$%
_{S}$ are found in the compounds with lower tolerance factors: Nd$_{0.5}$Ca$%
_{0.5}$MnO$_{3}$ and Ho$_{0.5}$Ca$_{0.5}$MnO$_{3}$. The tolerance factor
compares the Mn-O separation with the separation of the oxygen atom and
A-site occupant in AMnO$_{3}$. In this way the tolerance factor
characterizes the angle of the Mn-O-Mn bond. Compounds with higher tolerance
factors present wider conduction bands, which lead to a stronger
ferromagnetic interactions and lower electrical resistivity.

\section{Conclusions}

We have made a magnetic characterization of Nd$_{0.5}$Sr$_{0.5}$MnO$_{3}$, Nd%
$_{0.5}$Ca$_{0.5}$MnO$_{3}$ and Ho$_{0.5}$Ca$_{0.5}$MnO$_{3}$
polycrystalline samples. It allowed us to identify the ferromagnetic,
antiferromagnetic and charge ordering phases in each case. There are two
important differences among these samples. The first one is that the
tolerance factor, which influences the width of the conduction band,
decreases from Nd$_{0.5}$Sr$_{0.5}$MnO$_{3}$ to Ho$_{0.5}$Ca$_{0.5}$MnO$_{3}$%
. The second one is that the rare earth ions have intrinsic magnetic
moments: J=9/2 for Nd$^{3+}$ and J=8\ for Ho$^{3+}$. These intrinsic
magnetic moments experience a short range order at low temperatures.
Moreover, the relation between the charge ordering temperature and the
antiferromagnetic ordering temperature changes from one sample to the other.
In the Nd$_{0.5}$Sr$_{0.5}$MnO$_{3}$ sample they are approximately
coincident (T$_{CO}\approx $T$_{N}$), in the Nd$_{0.5}$Ca$_{0.5}$MnO$_{3}$
sample the charge ordering temperature is much higher (T$_{CO}\gg $T$_{N}$),
and in the Ho$_{0.5}$Ca$_{0.5}$MnO$_{3}$ sample there is no a long range
antiferromagnetic transition.

We also reported, to our knowledge for the first time, specific heat
measurements with applied magnetic fields between 0 and 9 T and temperatures
between 2 and 300 K in all these three samples. Each curve was successfully
fitted at high temperatures by an Einstein model with three optical phonon
modes. Close to the charge ordering and ferromagnetic transition
temperatures the specific heat curves showed peaks superposed to the
characteristic response of the lattice oscillations. The entropy variation
corresponding to the charge ordering transition was higher than the one
corresponding to the ferromagnetic transition. Near 160 K in the two
compounds with Nd$^{3+}$ ions the specific heat curve showed an abrupt
change in slope, which were correlated to the corresponding
antiferromagnetic transition.

Absolute values of specific heat close to 2 K were about 100 times higher in
our samples than in other charge ordering samples like Pr$_{0.63}$Ca$_{0.37}$%
MnO$_{3}$. At low temperatures the specific heat curve, in all three studied
samples and measured magnetic fields, showed a Schottky-like anomaly. In all
cases an increase in the applied magnetic field moves the Schottky peak to
higher temperatures. However, the height of the peak and the entropy, behave
differently in Nd$_{0.5}$Sr$_{0.5}$MnO$_{3}$ than in the other two samples.
We could not successfully fit the curves by either assuming a singlet or a
distribution of two-level-Schottky anomaly. Besides, we did not find a
straightforward correlation between the maximum temperature of the Schottky
anomaly and the magnetic moment of the rare earth ions. More experiments are
clearly necessary to unambiguously identify the origin of the Schottky
anomaly and its possible correlation with the charge ordered phase.

We thank the Brazilian science agencies FAPESP and CNPq for the financial
support.

\bigskip 

\newpage

\begin{center}
FIGURE CAPTIONS
\end{center}

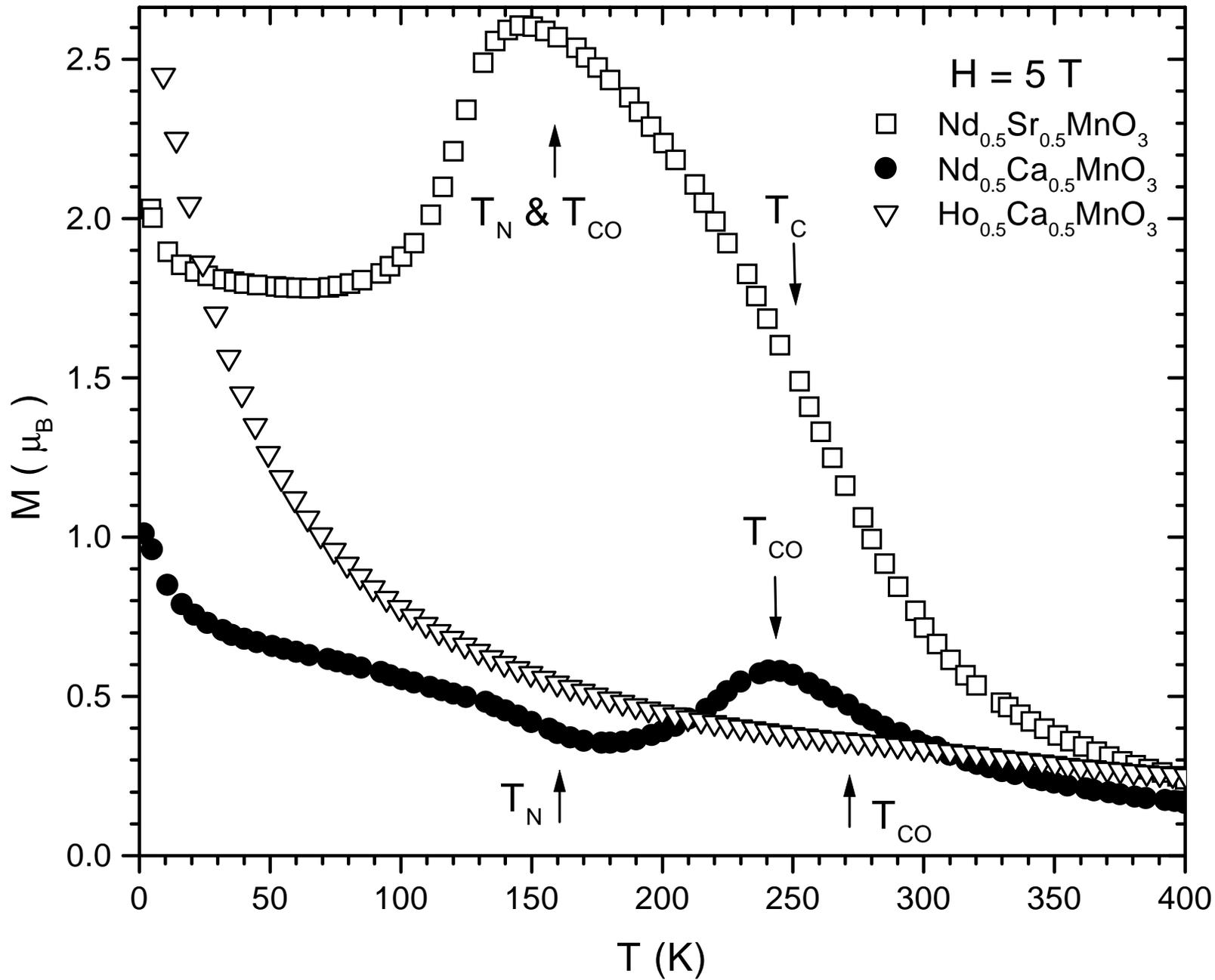
Figure 1. Temperature dependence of the magnetization, with a 5 T applied
magnetic field, in field cooling--warming condition for the three
polycrystalline samples studied. Magnetization is given in Bohr magnetons
per manganese ion. It is indicated T$_{C}$, T$_{N}$ and T$_{CO}$ for in each
sample. Note that charge ordering can occur in the presence of a
ferromagnetic, antiferromagnetic or\ a paramagnetic phase.

\bigskip 

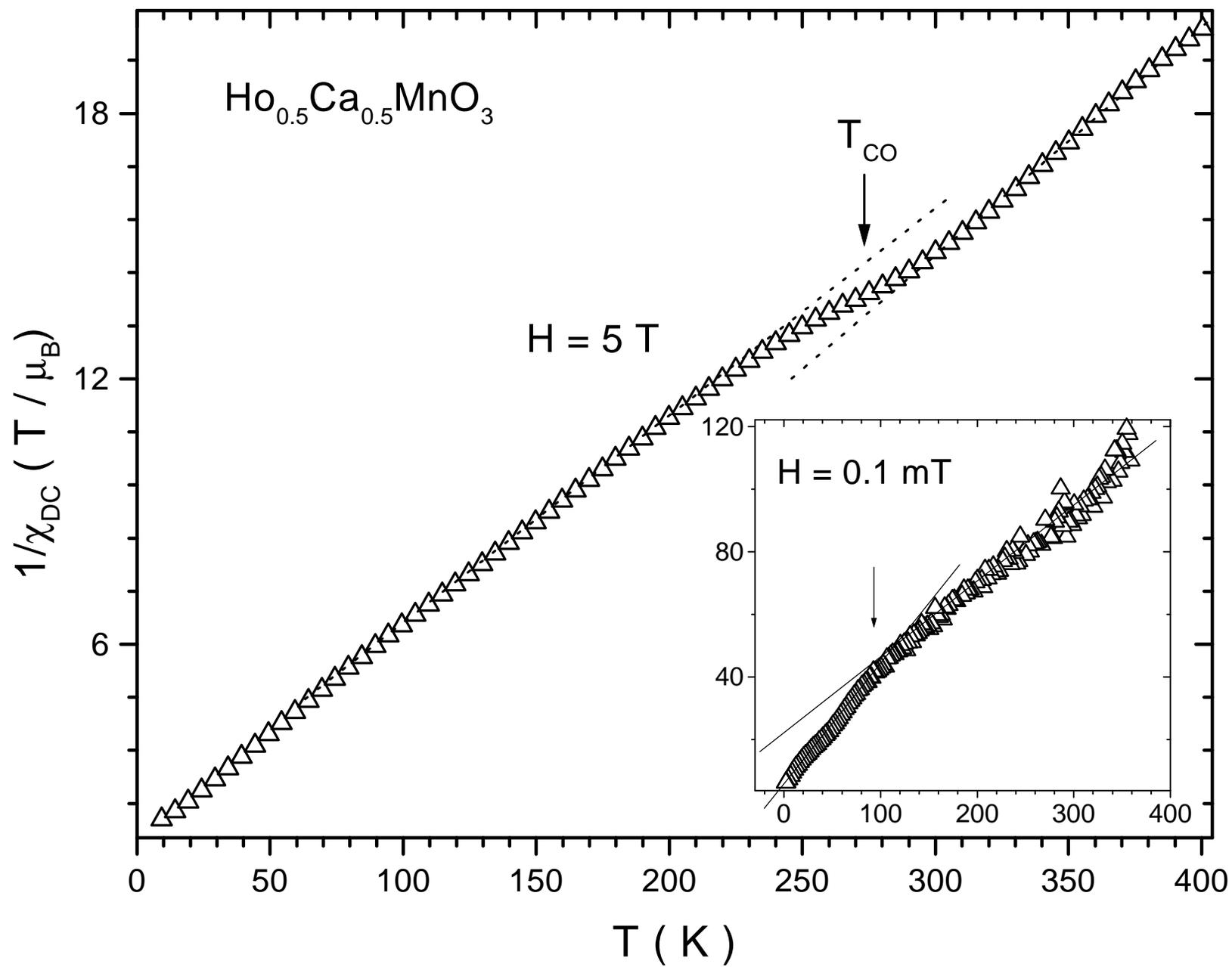
Figure 2. Inverse of DC susceptibility measurements as a function of
temperature in the Ho$_{0.5}$Ca$_{0.5}$MnO$_{3}$\ sample. The main graph
shows the measurement made with a 5 T applied magnetic field, while the
inset with a probe magnetic field of 0.1 mT. Lines are only guides to the
eye.

\bigskip 

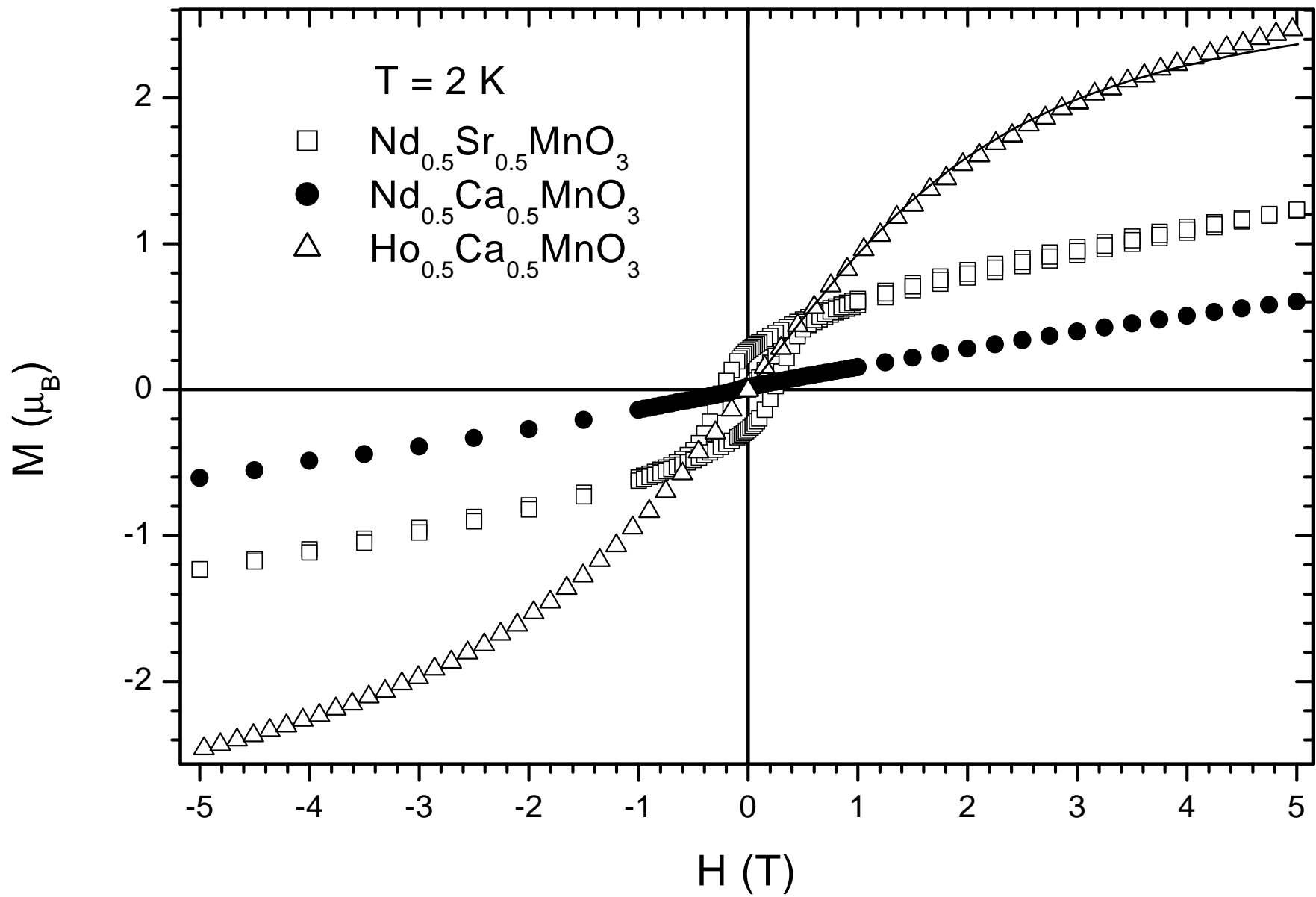
Figure 3. Magnetization versus applied magnetic field at 2 K for the three
samples studied. After a zero field cooling the magnetic field was increased
from 0 to 5 T, decreased from 5 T to -5 T and increased again from -5 T to 5
T. The solid line represents a fitting by a Brillouin function (See the text
for details). Magnetization is given in Bohr magnetons per manganese ion.

\bigskip 

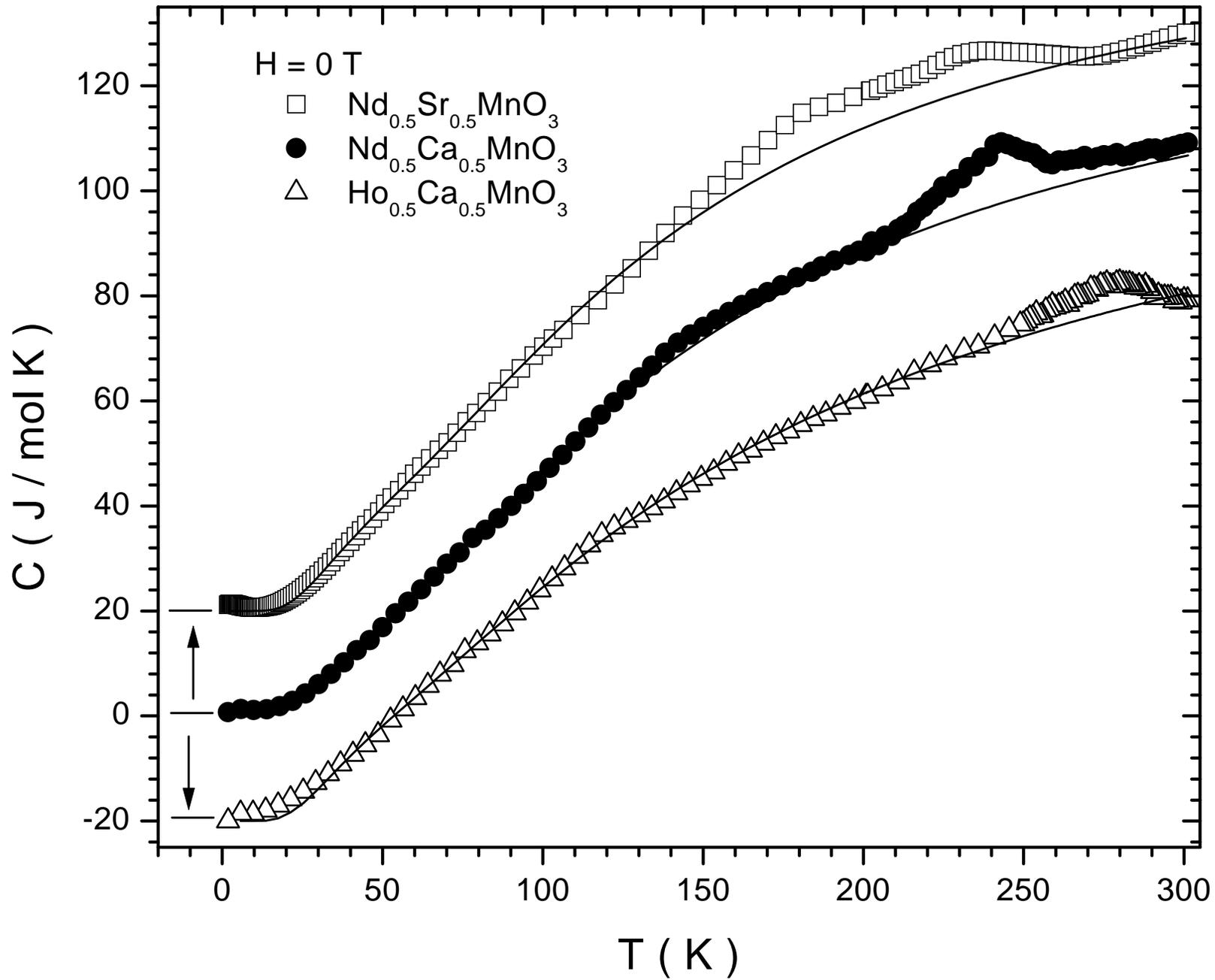
Figure 4. Specific heat measurements with zero magnetic field from 2 to 300
K for Nd$_{0.5}$Sr$_{0.5}$MnO$_{3}$ (open squares), Nd$_{0.5}$Ca$_{0.5}$MnO$%
_{3}$ (closed circles) and Ho$_{0.5}$Ca$_{0.5}$MnO$_{3}$\ (open triangles).
Continuous lines represent the fitting of the phonon background to the
Einstein model (see text). The Nd$_{0.5}$Sr$_{0.5}$MnO$_{3}$ and Ho$_{0.5}$Ca%
$_{0.5}$MnO$_{3}$ curves were displaced 20 J/mol K upside and downside,
respectively.

\bigskip 

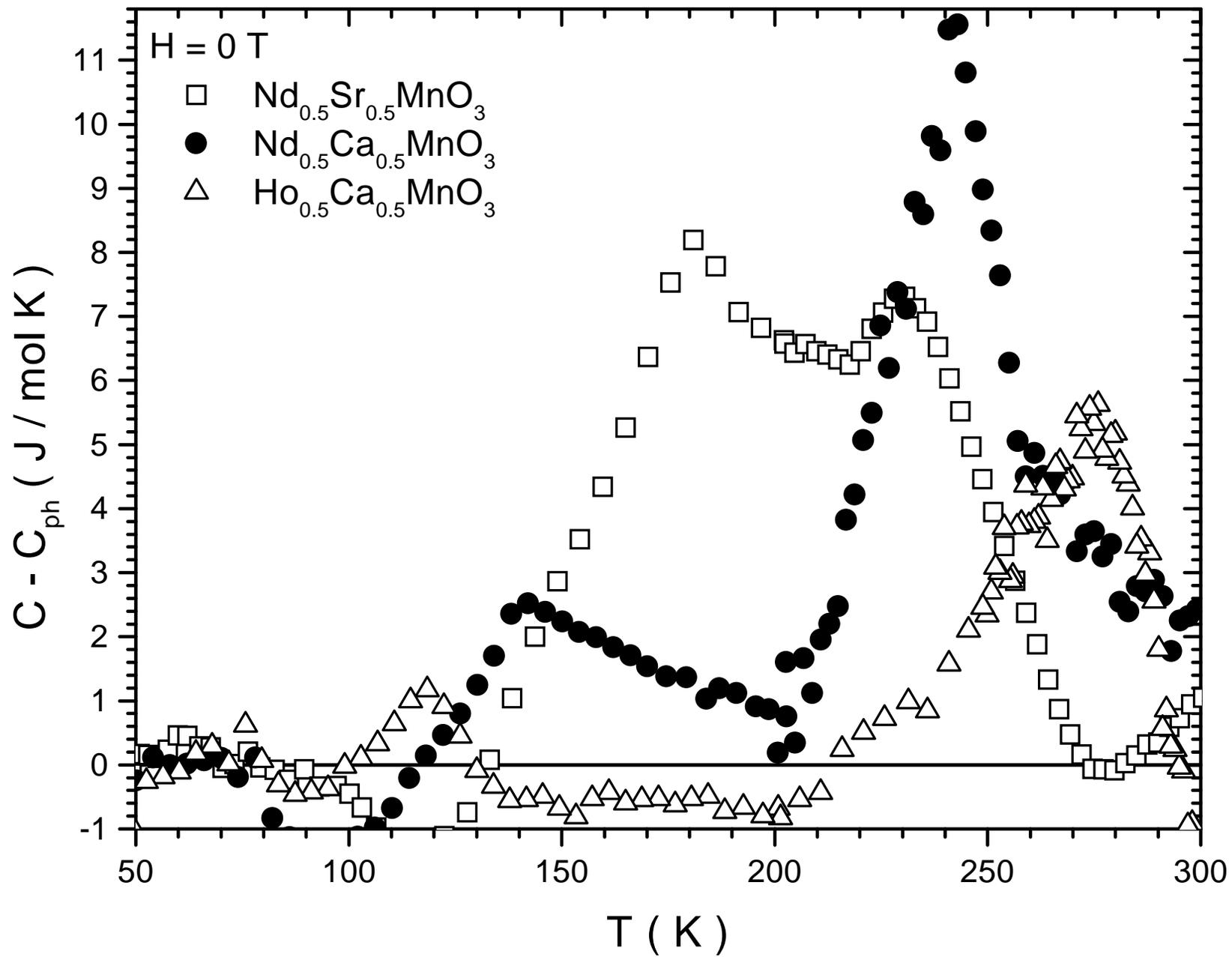
Figure 5. Differences between the experimental specific heat data and the
corresponding phonon background curves in the Nd$_{0.5}$Sr$_{0.5}$MnO$_{3}$
sample (open squares), Nd$_{0.5}$Ca$_{0.5}$MnO$_{3}$ sample (closed circles)
and Ho$_{0.5}$Ca$_{0.5}$MnO$_{3}$\ sample (open triangles).

\bigskip 

Figure 6. Specific heat measurements between 2 and 30 K in the samples of
(a) Nd$_{0.5}$Sr$_{0.5}$MnO$_{3}$, (b) Nd$_{0.5}$Ca$_{0.5}$MnO$_{3}$ and (c)
Ho$_{0.5}$Ca$_{0.5}$MnO$_{3}$. Measurements were made in the presence of
applied magnetic fields of 0 T (open squares), 5 T (closed circles), 7 T
(open up triangles) and 9 T (closed down triangles). Continuous lines
represent the fitting of the 15 to 30 K temperature interval data to the
phonon contribution, as it is explained in the text. The graphs in the
insets show the difference between the experimental values and the
extrapolation of the phonon contribution to temperatures lower than 15 K.

\bigskip 

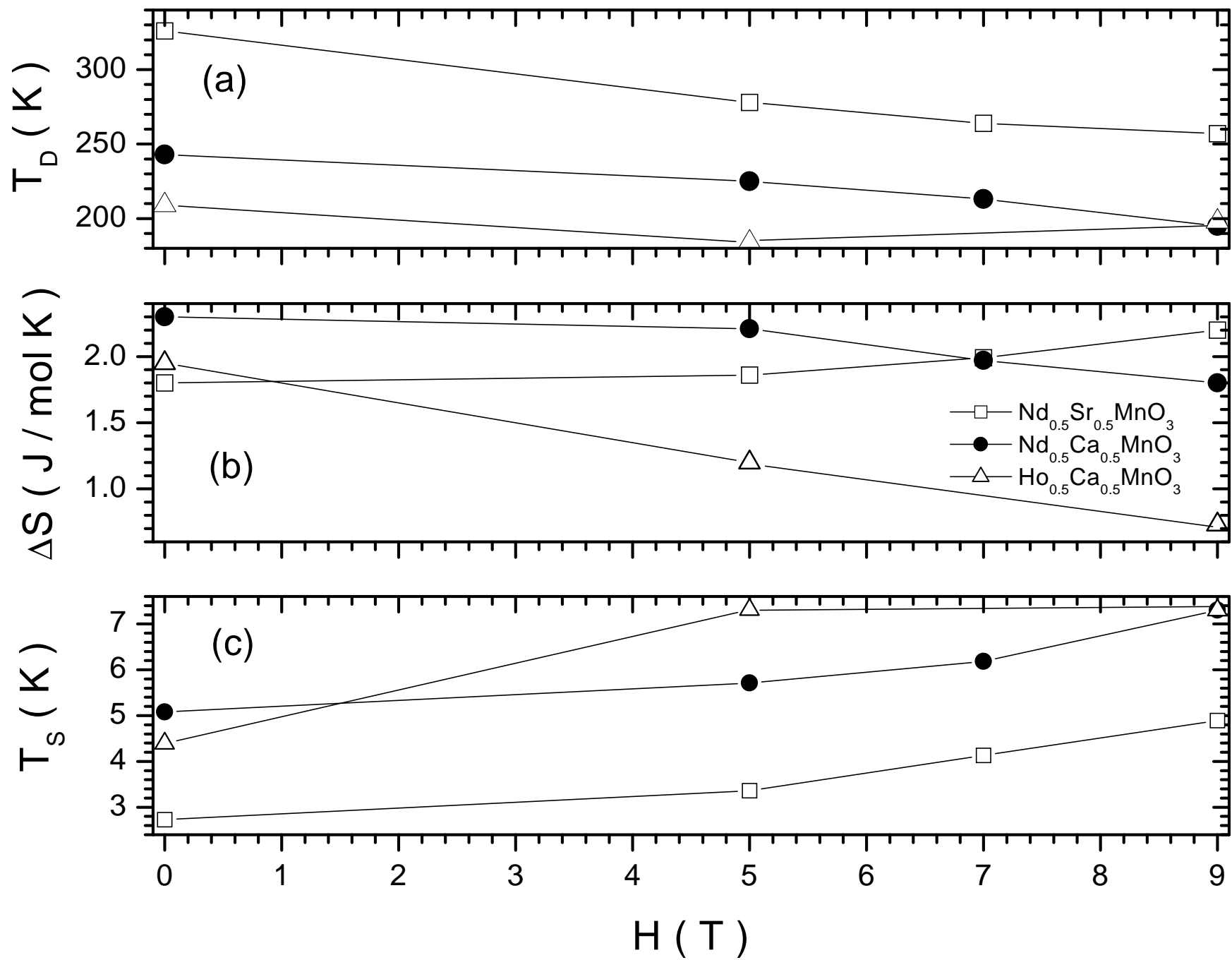
Figure 7. Debye temperature (T$_{D}$), entropy variation between 2 and 20 K (%
$\Delta S$) and Schottky temperature (T$_{S}$) as a function of the applied
magnetic field in Nd$_{0.5}$Sr$_{0.5}$MnO$_{3}$ (open squares), Nd$_{0.5}$Ca$%
_{0.5}$MnO$_{3}$ (closed circles) and Ho$_{0.5}$Ca$_{0.5}$MnO$_{3}$ (open
triangles).

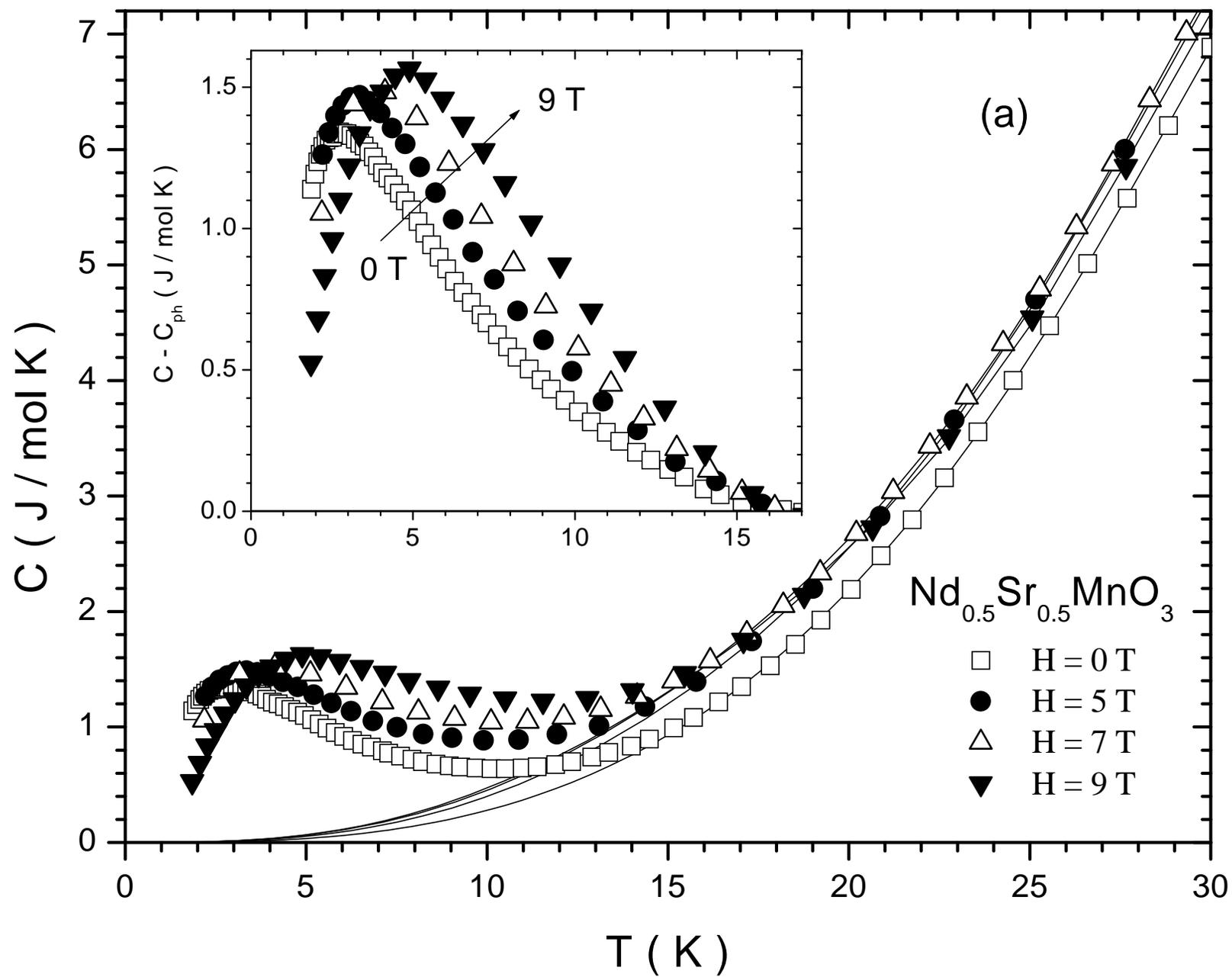
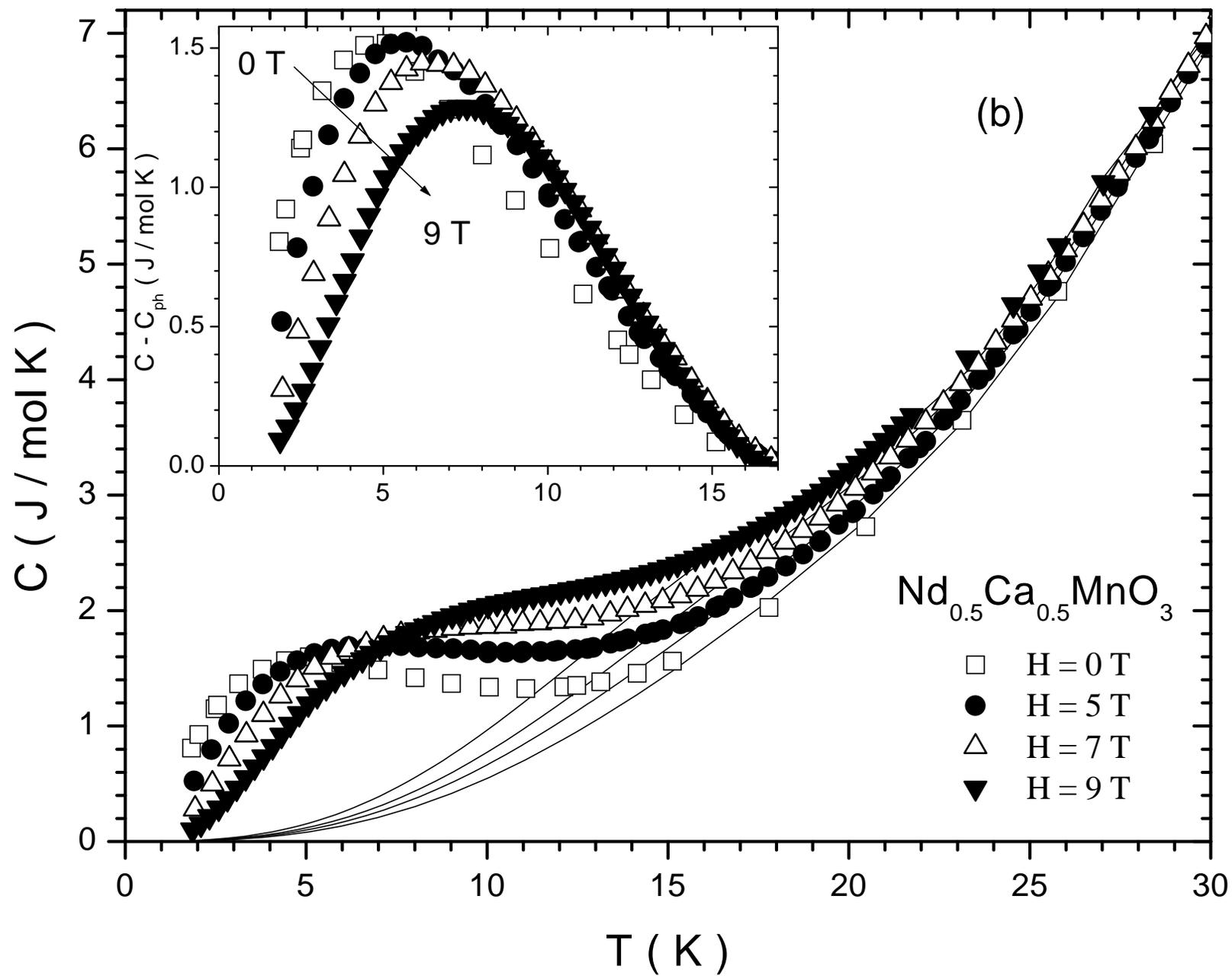
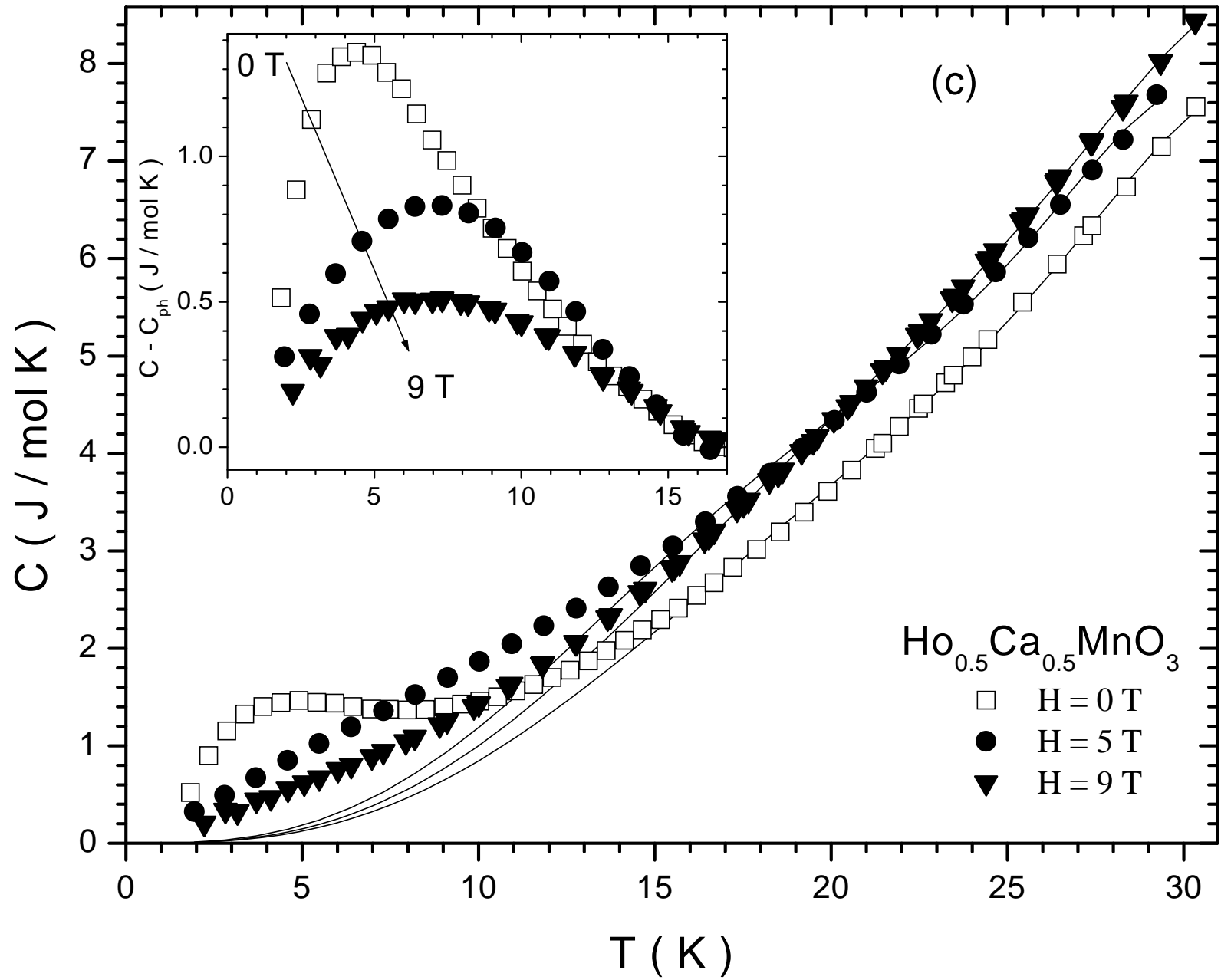
\end{document}